\documentclass[prl,twocolumn,preprintnumbers,amsmath,amssymb,floatfix]{revtex4}

\usepackage{graphicx}
\usepackage{epstopdf}
\usepackage{dcolumn}
\usepackage{bm}
\usepackage{color}
\usepackage{natbib}
\usepackage{amsmath}
\usepackage{amssymb}

\newcommand{\ct}{Ca$_3$Ti$_2$O$_7$}
\newcommand{\cm}{Ca$_3$Mn$_2$O$_7$}

\begin{document}

\title{Hybrid improper ferroelectricity: a mechanism for controllable polarization-magnetization coupling}
\author{Nicole A. Benedek  and Craig J. Fennie}
\affiliation{School of Applied $\&$ Engineering Physics, Cornell University, Ithaca, NY 14853 USA}

\begin{abstract}

First-principles calculations are presented for the layered perovskite Ca$_3$Mn$_2$O$_7$. 
The results reveal a rich set of coupled structural, magnetic and polar domains in which oxygen octahedron rotations induce ferroelectricity, magnetoelectricity and weak-ferromagnetism.
The key point is that the rotation distortion is a combination of two non-polar modes with different symmetries. We use the term ``hybrid'' improper ferroelectricity to describe this phenomenon and discuss how control over magnetism is achieved through these functional antiferrodistortive octahedron rotations.
\end{abstract}
\pacs{75.85.+t, 77.80.-e,81.05.Zx }

\maketitle

The utility of multiferroics for low-power electronic devices stems from the possibility for electric-field control of magnetism at room temperature~\cite{Tokunaga:2009, Bibes:2008, Ramesh:2008,scott06}.  A challenge that has so far not been overcome is to identify a stable, single phase multiferroic material in which the magnetization can be deterministically switched 180$^{\circ}$. A large electrical polarization strongly coupled to the magnetization is generally thought to be a key requirement~\cite{scott06}. 

Magnetically-driven improper ferroelectrics, such as  TbMnO$_3$, are materials in which a spontaneous polarization arises due to symmetry-breaking by a spin instability~\cite{tokura03,cheong04}. These materials naturally have a strong coupling between magnetism and the polarization, but the polarization is too small for device applications.  In known multiferroic materials with a large electrical polarization, such as BiFeO$_3$~\cite{Ramesh:2003}, the ferroelectricity is proper, originating from a zone-center polar lattice instability, as in the prototypical perovskite ferroelectric PbTiO$_3$. However, except in a few special cases that satisfy restrictive symmetry criteria, the polar instability in a proper ferroelectric does not break the right symmetries to turn on a nonzero magnetization and therefore does not satisfy the criteria of the only known mechanism that enables the electric field switching of the magnetization~\cite{FennieFeTiO}.

It is desirable to identify a more general mechanism -- applicable to a large class of materials, for example, the ABO$_3$ perovskites -- whereby ferroelectricity and ferromagnetism are induced by the same lattice instability. Octahedron rotations, ubiquitous in perovskites and related materials, are natural candidates for this lattice instability as they are known to strongly couple to magnetic properties~\cite{Millis, Goto,Rini:2007}.  Unfortunately, such distortions (or combinations of distortions) in simple perovskites are not polar and therefore do not induce ferroelectricity.
Recently, however, Bousquet, \textit{et al.}~\cite{bousquet08} made a key discovery that by layering perovskites in an artificial superlattice, e.g., (SrTiO$_3$)/(PbTiO$_3$), a polarization can arise from the coupling of two rotational modes.  Taking advantage of this mechanism to realize a strongly coupled multiferroic is a challenge.

In this Letter, we demonstrate how octahedron rotations simultaneously induce ferroelectricity,  magnetoelectricity, and weak ferromagnetism in a class of naturally occurring (ABO$_3$)$_2$(AO) layered perovskites. 
The key point is that the polarization, $P$, arises from a rotation pattern that is a combination of two non-polar lattice modes with different symmetries, $P \sim \mathcal{R}_1 \mathcal{R}_2$, as in Ref.~\onlinecite{bousquet08}, but here rotations $\mathcal{R}_1$ and $\mathcal{R}_2$ additionally induce magnetoelectricity and weak ferromagnetism respectively.
We use the term ``hybrid'' improper ferroelectricity to describe this ferroelectric mechanism (in loose analogy to improper ferroelectricity~\cite{Levanyuk} such as in YMnO$_3$~\cite{FennieYMnO}) in order to generalize the idea to include cases where the two distortion patterns do not necessarily  condense at the same temperature.
This mechanism has no impediment to room temperature operation and in fact opens up entirely new classes of materials in which to search for strongly-coupled multiferroics. 
Our results show a rich set of coupled structural, magnetic and polar domains and suggest the possibility to switch between magnetic domains with an electric field.

%

%
We have identified (CaBO$_3$)$_2$CaO, with B=Ti~\cite{white91}, Mn~\cite{martin01,Lobanov}, as two materials that display hybrid improper ferroelectricity.  It is significant that they occur in nature in bulk,  forming in the Ruddlesden-Popper (RP) homologous family with general formula A$_{n+1}$B$_n$O$_{3n+1}$. 
Any given member of the RP series consists of ABO$_3$ perovskite blocks stacked along the [001] direction with an extra AO sheet inserted every $n$ perovskite unit cells. 
For  \cm~($n=2$) the experimental picture of the sequence of phase transitions from the paraelectric $I4/mmm$ phase to the ferroelectric $A2_1am$ phase is not clear. Two possibilities have been proposed: (1) $I4/mmm \rightarrow Cmcm \rightarrow A2_1am$, and (2) a direct transition from $I4/mmm \rightarrow  A2_1am$.
Additionally, it has been shown to display weak ferromagnetism~\cite{Jung,Lobanov}.
To our knowledge, \ct~ has only been reported in the polar $A2_1am$ structure.
In the remainder of this Letter, we focus mainly on the magnetic compound,  \cm, as a prototype of this class of strongly coupled multiferroics.

First-principles calculations were performed using DFT using PAW potentials within LSDA+U~\cite{anisimov97} as implemented in {\sf VASP}~\cite{VASP1,VASP2,PAW1, PAW2}. All calculations were repeated with the PBEsol functional, which provides an improved description of structural parameters; there was no qualitative change in any of our results.  We used U = 4.5 eV and  J$_{\rm H}$=1 eV for the Mn-ion on-site Coulomb and exchange parameters respectively. Where noted, non-collinear calculations with L-S coupling were performed. We used a 600eV plane wave cutoff, a 4$\times$4$\times$2 Monkhorst-Pack mesh, which we checked for sufficient accuracy.  


%
%
%

\begin{figure}
\centering
\includegraphics[width=6cm]{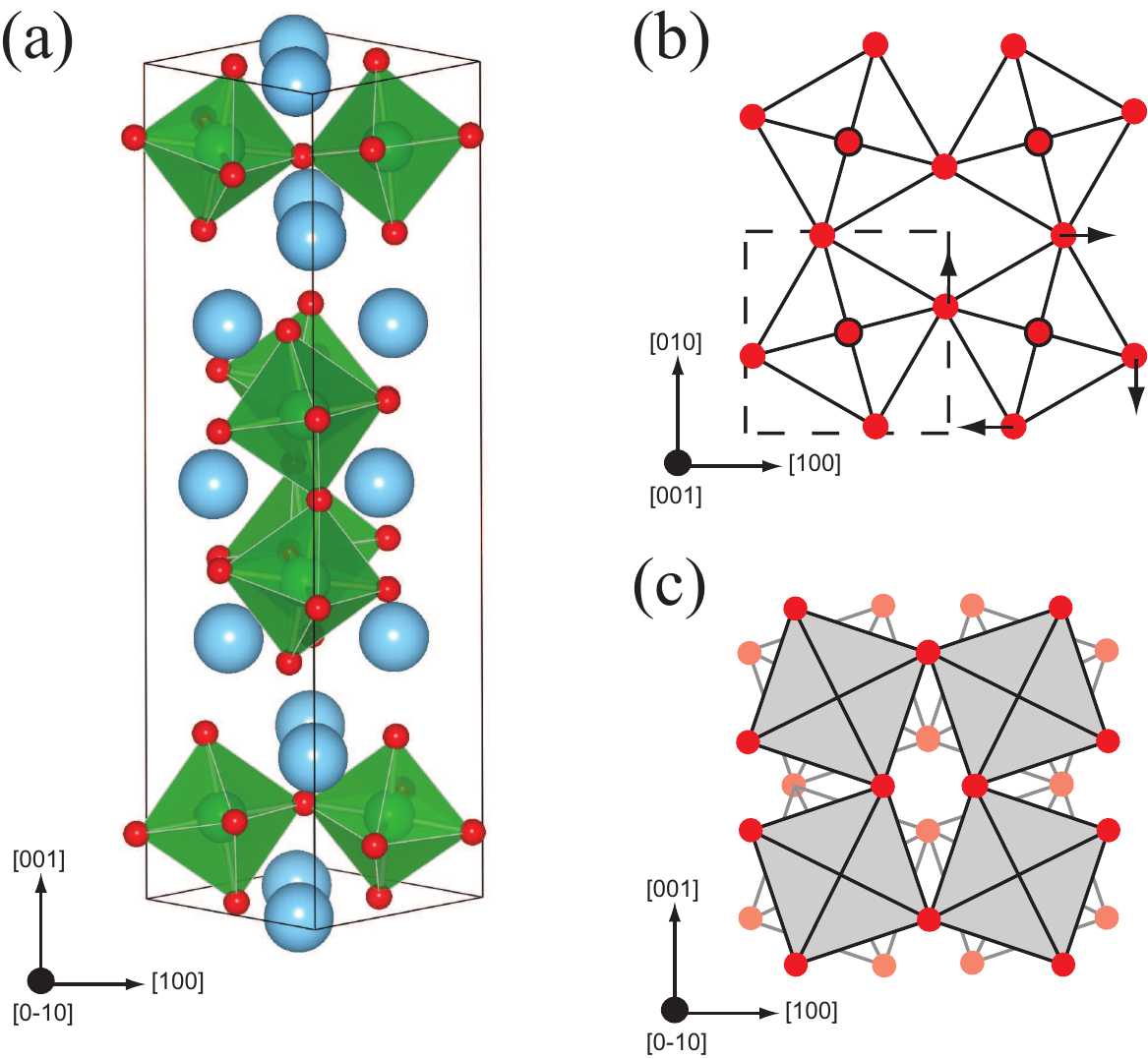}
\caption{\label{structures}{Ca$_3$Mn$_2$O$_7$ structure and rotation distortions.} 
\textbf{a}, The $A2_1am$ ferroelectric ground state structure. Blue (large) spheres correspond to Ca ions. 
\textbf{b} Schematic of the atomic displacements corresponding to the $X_2^+$ rotation. The dashed square denotes the unit cell of the $I4/mmm$ parent structure. \textbf{c}, Schematic of the  $X_3^-$ tilt mode. All axes refer to the coordinate system of the $I4/mmm$ parent structure.}
\end{figure}

In the polar $A2_1am$ structure, the oxygen octahedra are significantly rotated and tilted with respect to the  $I4/mmm$ structure, as shown in Figure \ref{structures}. The polarization, in the $xy-$plane by symmetry, is found from our first-principles calculations to be large, $P\approx$ 5 $\mu$C/cm$^2$ ($P\approx$ 20 $\mu$C/cm$^2$ for \ct).
Group theoretical methods show that $A2_1am$ is related to $I4/mmm$ by three distinct atomic distortions: a polar zone-center mode transforming like the irreducible representation (irrep) $\Gamma_5^-$, and two zone-boundary modes at the $X$ (1/2, 1/2, 0) point  -- an oxygen octahedron rotation mode with irrep $X_2^+$ and an oxygen octahedron tilt mode with irrep $X_3^-$.  Note that $X_2^+ \oplus X_3^-$ establishes the $A2_1am$ space group, a zone-center polar instability is not required. Hence, it is possible to reach the ferroelectric state by means of a combination of rotations and tilts only. 

\begin{figure}
\includegraphics[width=7cm]{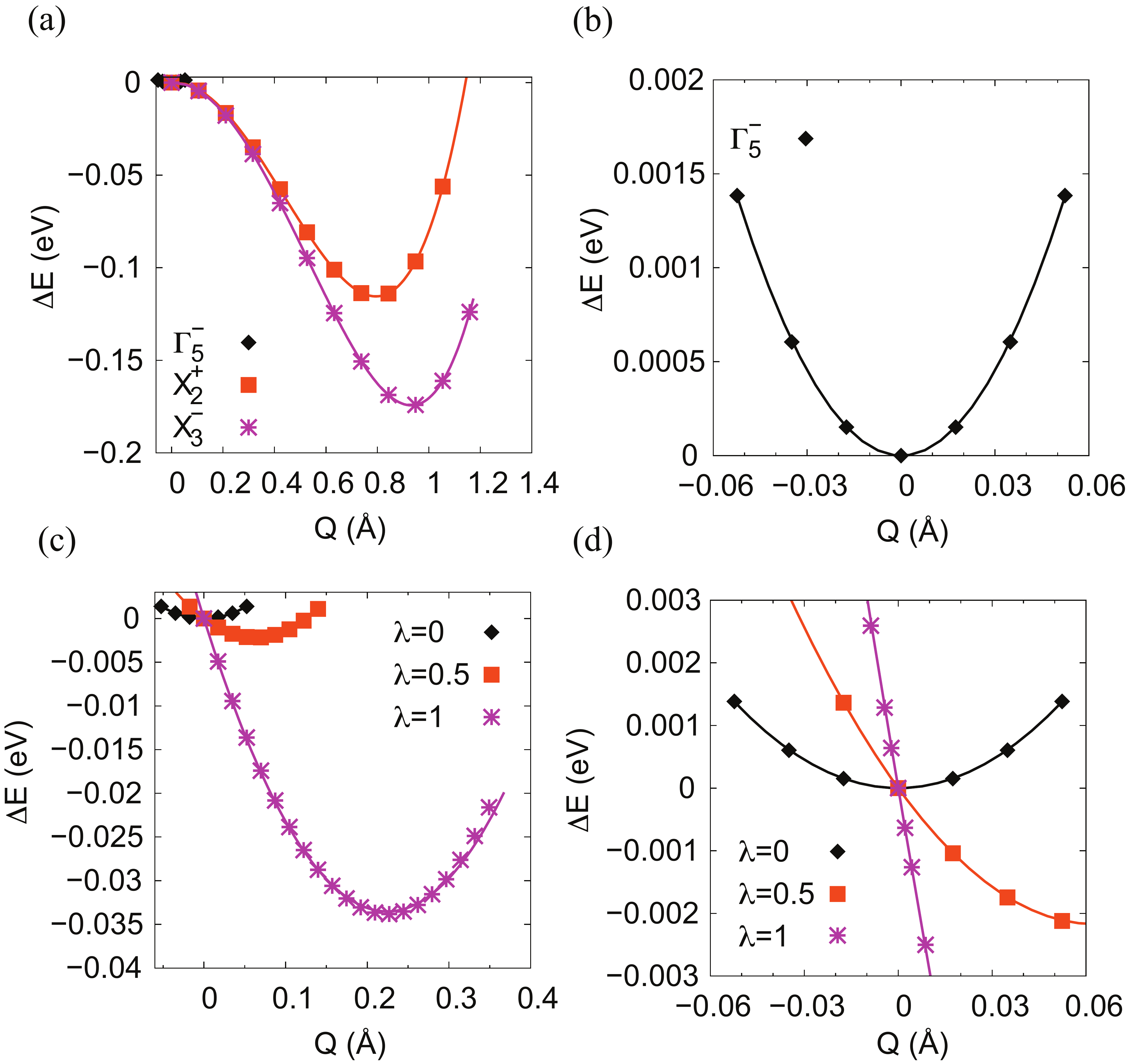}
\caption{\label{coupling} Energy surface about paraelectric $I4/mmm$ in \cm. Change in energy per formula unit as a function of the 
amplitude of \textbf{a}, the $X_2^+$ tilt and $X_3^-$ rotation modes and 
\textbf{b}, the polar $\Gamma_5^-$ distortion. \textbf{c} and 
\textbf{d}, Polarization in the presence of a hybrid order parameter, 
$Q_{X_{23}}$=$\lambda Q_{X_3^-}Q_{X_2^+}$. Note the differences in scales between panels.} 
\end{figure}

We project out the contribution of each $X_3^-$,  $X_2^+$, and  $\Gamma_5^-$ mode to the $A2_1am$ ground state structure and calculate from first principles the $T=0$ energy surface around the $I4/mmm$ reference structure.
Figure \ref{coupling}a shows the total energy as a function of the  amplitude of the distortion for the individual rotation ($Q_{X_2^+}$),  tilt ($Q_{X_3^-}$), and polar ($Q_{\Gamma_5^-}$) distortions. 
Relatively large energy gains can be seen within a characteristic double-well potential for the rotation and tilt distortions, whereas the polar contribution is stable, as shown in Figure~\ref{coupling}b.
Additionally, Figure 3 shows that the combination of $Q_{X_2^+}$ plus $Q_{X_3^-}$ lowers the energy -- even in the absence of the $Q_{\Gamma_5^-}$ distortion -- resulting in a ground state with four structural domains.

Figure \ref{coupling}c,d shows how the polarization arises from a coupling to  a ``hybrid'' order parameter $Q_{X_{23}} =  Q_{X_2^+}Q_{X_3^-}$. In the absence of rotation and tilt distortions ($Q_{X_{23}}$=0), the polarization has a single minimum at $P$ =0. As $Q_{X_{23}}$ increases, the polarization never becomes unstable. Rather, the minimum shifts  to a non-zero value. The result of increasing $Q_{X_{23}}$ is analogous to the effect of turning on a finite electric field, just like in the classic case of improper ferroelectricity~\cite{Levanyuk}. Furthermore, when $Q_{X_{23}} \ne 0$, the polarization  is linear about zero (Figure \ref{coupling}d), a direct indication of improper coupling $$\mathcal{F} = \alpha P Q_{X^+_2}Q_{X^-_3}$$ between the polarization, rotations, and tilts. 

 These observations suggest that the rotation mode and the tilt mode are the primary modes driving the transition to the ferroelectric $A2_1am$ phase. This single distortion pattern, $Q_{X_{23}}$, is the hybrid improper mode. 
In contrast to proper and conventional improper ferroelectrics, more than one lattice distortion may switch the polarization in a hybrid improper ferroelectric. Symmetry implies, and our  calculations confirm, that the polarization reverses by either switching a $X^+_2$ distortion or a $X^-_3$ distortion but not both, resulting in two polar domains.
 
%
%

As is often the case in perovskites and related materials, octahedron rotations directly couple to the magnetic ordering. 
We determined from first principles the magnetic ground state of $A2_1am$  \cm~with polarization along [010] to be antiferromagnetic (G-type within the perovskite bilayer). The spins point along [001] due to crystalline anisotropy. Additional spin-orbit interactions give rise to a net spin-canted moment of $M\approx0.18 \mu_B$ per unit cell (4 spins) along [100]. These results are consistent with previous experiments.  Note that the magnetic point group, $2'mm'$, allows for a linear magnetoelectric effect~\cite{dzy2}, which symmetry indicates is induced by the $X_2^+$ rotation distortion.
Application of Dzyaloshinskii's criteria~\cite{dzy1,Moriya} shows the canted moment is the result of the $X^-_3$  tilt distortion. Indeed, if we compute the magnetic ordering in the $Cmcm$ ($Cmca$) structure obtained by freezing in the $X^-_3$ ( $X^+_2$) mode alone, we find $M\approx0.22 \mu_B$  ($M = 0\mu_B$) per unit cell, which reverses with reversal of this octahedral tilt~\cite{EPAPS} resulting in two magnetic domains.

\begin{figure}
\includegraphics[width=8cm]{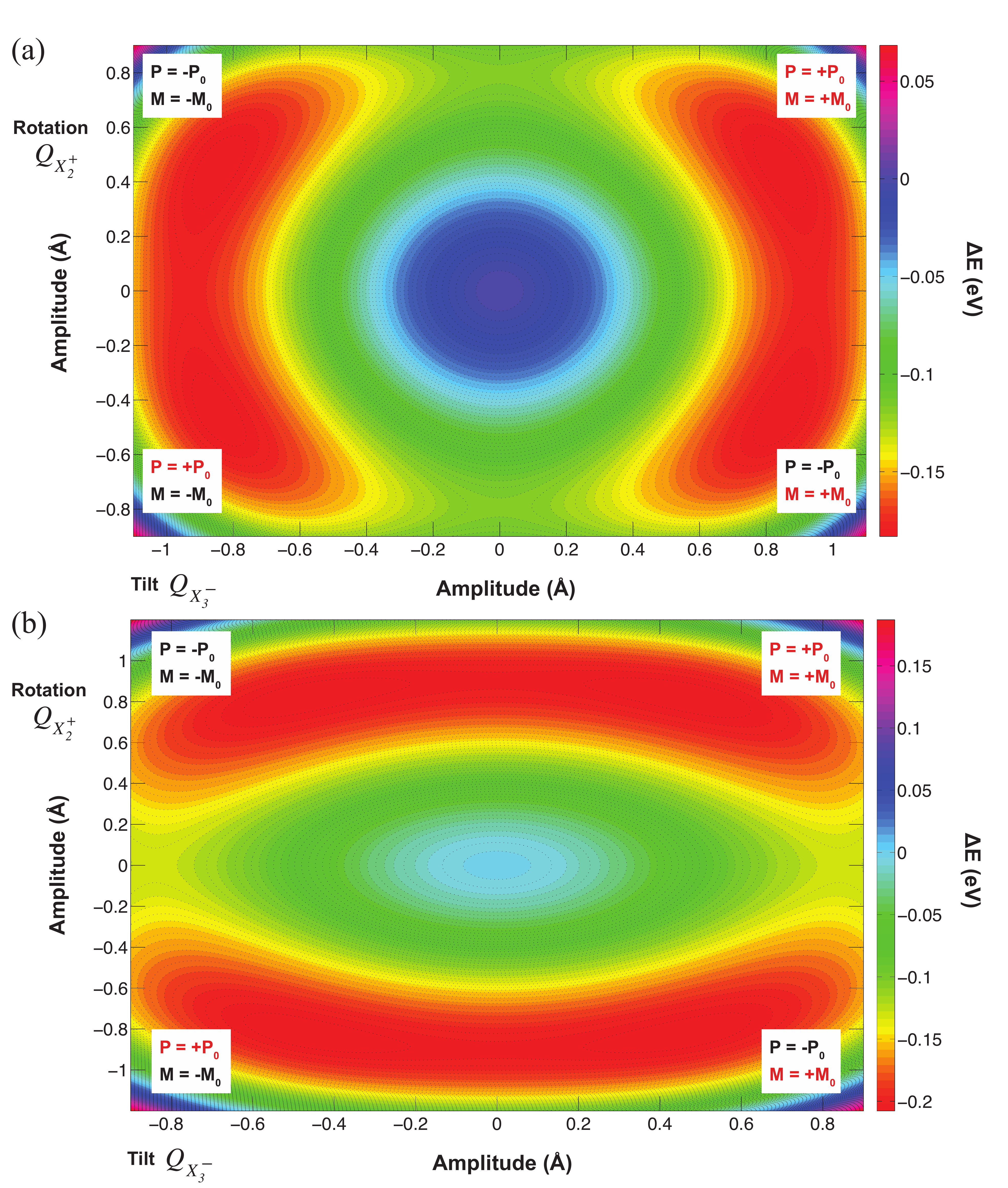} 
\caption{\label{domains}{Four types of structural domains and possible paths for electric-field switching of the magnetization in \cm.} \textbf{a}, Unstrained (bulk): Starting from a fixed $X^-_3$ tilt domain, switch polarization by switching $X^+_2$ rotation: $P_0\rightarrow -P_0$ and $M_0\rightarrow M_0$. \textbf{b}, Compressive biaxial strain: Start from a fixed $X^+_2$ rotation domain, switch polarization by switching $X^-_3$ tilts: $P_0\rightarrow -P_0$ and $M_0\rightarrow -M_0$. The energy change is per formula unit.}
\end{figure}

So which distortion, the rotation or the tilt,  reverses in an electric-field switching experiment? Although polarization switching in a ferroelectric is a complex, dynamic process, we may gain some insight by examining the intrinsic energy barriers between domains.
As shown in Figure~\ref{domains}a the lowest energy pathway to switch the direction of the polarization is along the $X^+_2$ switching path. In this process the magnetization does not reverse its sign.
The linear magnetoelectric effect, however, is induced by the  $X^+_2$  distortion as mentioned. This electric-field tunable oxygen rotation distortion may lead to an enhanced magnetoelectric effect. Such calculations are beyond the scope of this Letter, but future theoretical and experimental studies should make this clear.  
 %

%
Oxygen rotations in perovskites are known to respond strongly to pressure and epitaxial strain. 
Figure~\ref{domains}b shows the energy landscape around the $I4/mmm$ paraelectric structure at 1.5$\%$ compressive strain. Now the lowest energy pathway to switch the polarization to a symmetry equivalent state is along the $X^-_3$ switching path, which as previously discussed, switches the direction of the spin-canted moment.
Therefore, for an epitaxial thin film compressively strained in the $A2_1am$ phase, we predict that switching the direction of the polarization with an electric field will switch the direction of the equilibrium magnetization by 180$^{\mathrm{o}}$. As in all experiments to date based on the linear magnetoelectric effect, however, a single antiferromagnetic domain must be annealed and maintained throughout the experiment. 

\begin{figure}
\includegraphics[width=8.5cm]{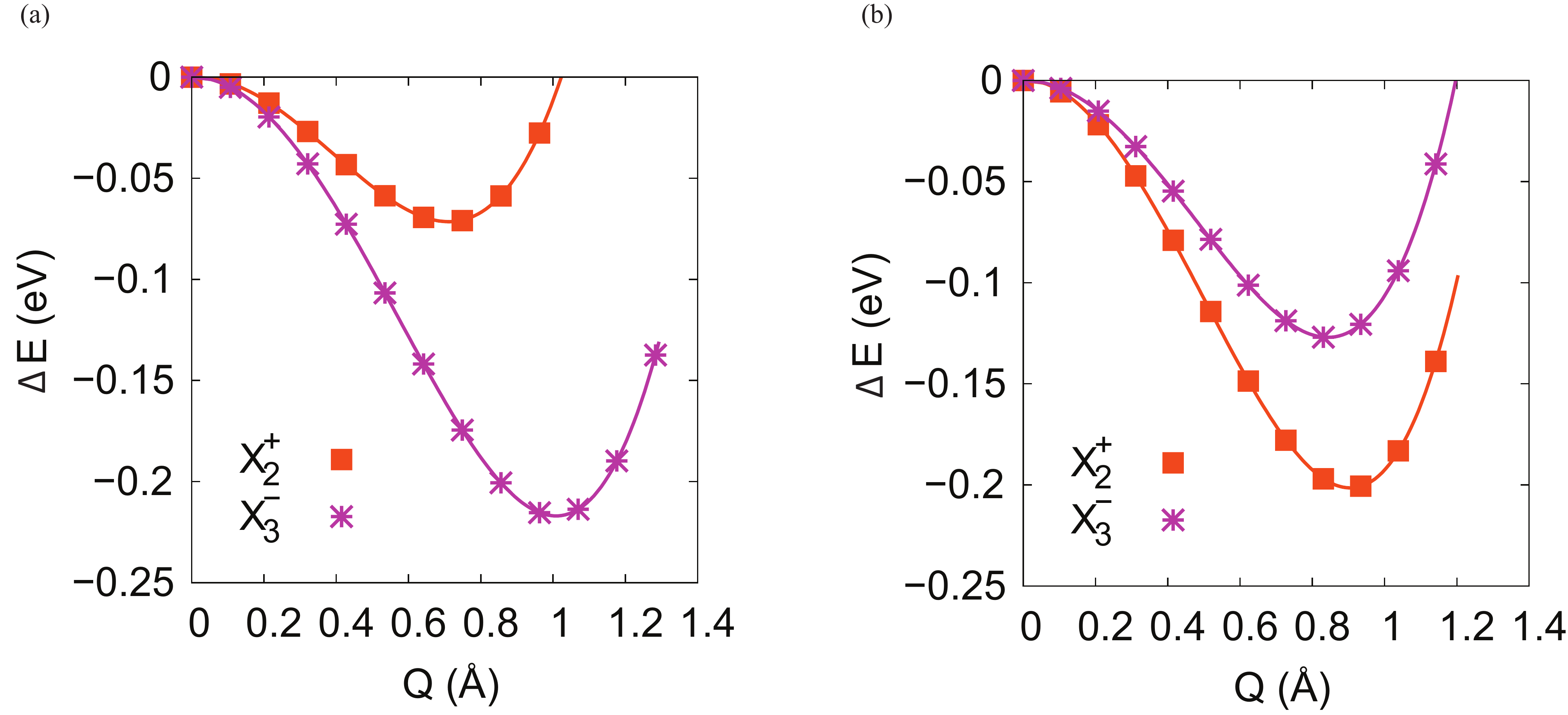}
\caption{\label{strain}{Effect of epitaxial strain on the unstable zone-boundary modes in Ca$_3$Mn$_2$O$_7$.} Change in energy per formula unit $w.r.t.$ paraelectric $I4/mmm$, as a function of the amplitudes of the $X_3^-$ and $X_2^+$ modes under  \textbf{a}, 1.5\% tensile  and \textbf{b}, 1.5\% compressive biaxial strain. The lines are a guide to the eye (similar results were also found for Ca$_3$Ti$_2$O$_7$). }
\end{figure}

This observation of tuning the intrinsic energy barriers between the domains with strain can be understood from well-known simple physical considerations~\cite{Samara}. Figure \ref{strain} shows the behavior of  the rotation and tilt distortions in \cm~under 1.5\% biaxial tensile and compressive strains. Figure \ref{strain}a shows that the energy lowering of the $X_2^+$ rotation is strongly reduced under tensile strain compared with the unstrained state shown in Figure \ref{coupling}a. 
Under compressive strain, the opposite behavior occurs: the $X_2^+$ mode is strongly favored, lowering the energy even more than the $X_3^-$ tilt mode, as shown in Figure \ref{strain}b. 
We emphasize that strain does not induce, enhance or mediate the polarization-magnetization coupling responsible for the electric-field switching of the magnetization in these materials. In the specific case of Ca$_3$Mn$_2$O$_7$,  strain merely alters the energy landscape around the $I4/mmm$ paraelectric structure, biasing one hybrid improper ferroelectric switching pathway (the one that simultaneously switches the direction of the magnetization) over the other. However, there is no fundamental reason why electric-field switching of the magnetization cannot be observed in bulk hybrid improper ferroelectrics.
%

%
%
In addition to the measurements already suggested, spatially resolving the structural, polar, and magnetic domains, e.g$.$, optically~\cite{fiebig},  should prove the coupling physics discussed even in bulk \cm.
It would also be of interest to understand the phase transition sequence from the high-symmetry paraelectric  $I4/mmm$ phase to the low-symmetry ferroelectric $A2_1am$ phase, which is impossible to prove or disprove from $T$=0 calculations.
In bulk, two possibilities have been proposed as we previously discussed~\cite{zzzStrainSeq}.
Path (1), having an intermediate $Cmcm$ phase, is consistent with Landau theory and with our calculated hierarchy of structural distortions as displayed in Figure~\ref{domains}a.
 If Path (2) turns out to be correct, however, this indicates that the two distortions making up the hybrid order parameter condense at the same temperature, exactly analogous to the recent discovery of Ref.~\onlinecite{bousquet08}. 
It is unlikely to be a coincidence that both PbTiO$_3$/SrTiO$_3$~\cite{bousquet08} and  (CaMnO$_3$)$_2$/CaO are layered perovskites. 
Regardless of the actual path taken -- or even if a paraelectric structure is realizable in the phase diagram of the system -- our conclusions on the coupling of rotation/tilt distortions to ferroelectricity and magnetism remain unchanged.

Note that the temperature scale of the hybrid improper ferroelectric mechanism is set by the structural distortions, which commonly occur above room temperature ($\sim$500-600K in \cm~\cite{bendersky03}). The `limiting' temperature in the case of Ca$_3$Mn$_2$O$_7$ is the N\'{e}el temperature, $T_N$ $\sim$115K \cite{Jung,Lobanov} and as such this prototype material is not ideal. There is, however, no fundamental reason why one couldn't discover (or design) a hybrid improper ferroelectric with a N\'{e}el temperature above room temperature (ongoing investigations into the connection between layering and hybrid improper ferroelectricity have realized initial materials design rules~\cite{Rondin}), so the mechanism has no impediment to room-temperature operation. 
This approach therefore shifts the challenge of discovering a  multiferroic in which the magnetization can be controlled by the electrical polarization  to the more familiar problem of designing a room temperature antiferromagnet.

In summary, we have introduced the term hybrid improper ferroelectricity to describe a state in which the polarization is {\it induced} by a complex distortion pattern consisting of more than one octahedron rotation mode. 
We have shown how this mechanism opens up new avenues to pursue strong polarization-magnetization coupling, as alluded to in Ref$.$ \onlinecite{bousquet08} and realized in this Letter, and discovered a new class of materials.
 We hope our Letter inspires further work in this intriguing field of materials with functional oxygen rotations and tilt distortions.\\

We acknowledge discussions with V$.$ Gopalan, D.G$.$ Schlom, K.M$.$ Rabe, and M$.$ Stengel. NAB was supported by the Cornell Center for Materials Research with funding from NSF  MRSEC program, cooperative agreement DMR 0520404. CJF was supported by the DOE-BES under Award Number DE-SCOO02334.\\

\section{Auxiliary Material}

\textbf{Landau theory to describe the polarization--magnetization coupling depicted in Figure 3.}\\
Note that in the main text we use the tetragonal, $I4/mmm$ structure (which has one formula unit per cell) as the paraelectric reference structure to discuss the hybrid order parameter. To make clearer the polarization-magnetization coupling, in this auxiliary material we use two different orthorhombic reference structures (which each have two formula units per cell) corresponding to the two different possible saddle points shown in Figure 3 of the main text.  Therefore due to the subsequent zone-folding, the atomic distortions that make up the order parameter that we refer to below as the polarization,  $\mathcal{P}$,  differ from those that make up the polar order parameter, $P$, from $I4/mmm$ defined in the main text.  This does not change the symmetry of the coupling of magnetization to the physical polarization, and hence does not change the conclusions that octahedron rotations induce ferroelectricity, magnetoelectricity and weak-ferromagnetism.

\begin{figure}
\centering
\includegraphics[width=8cm]{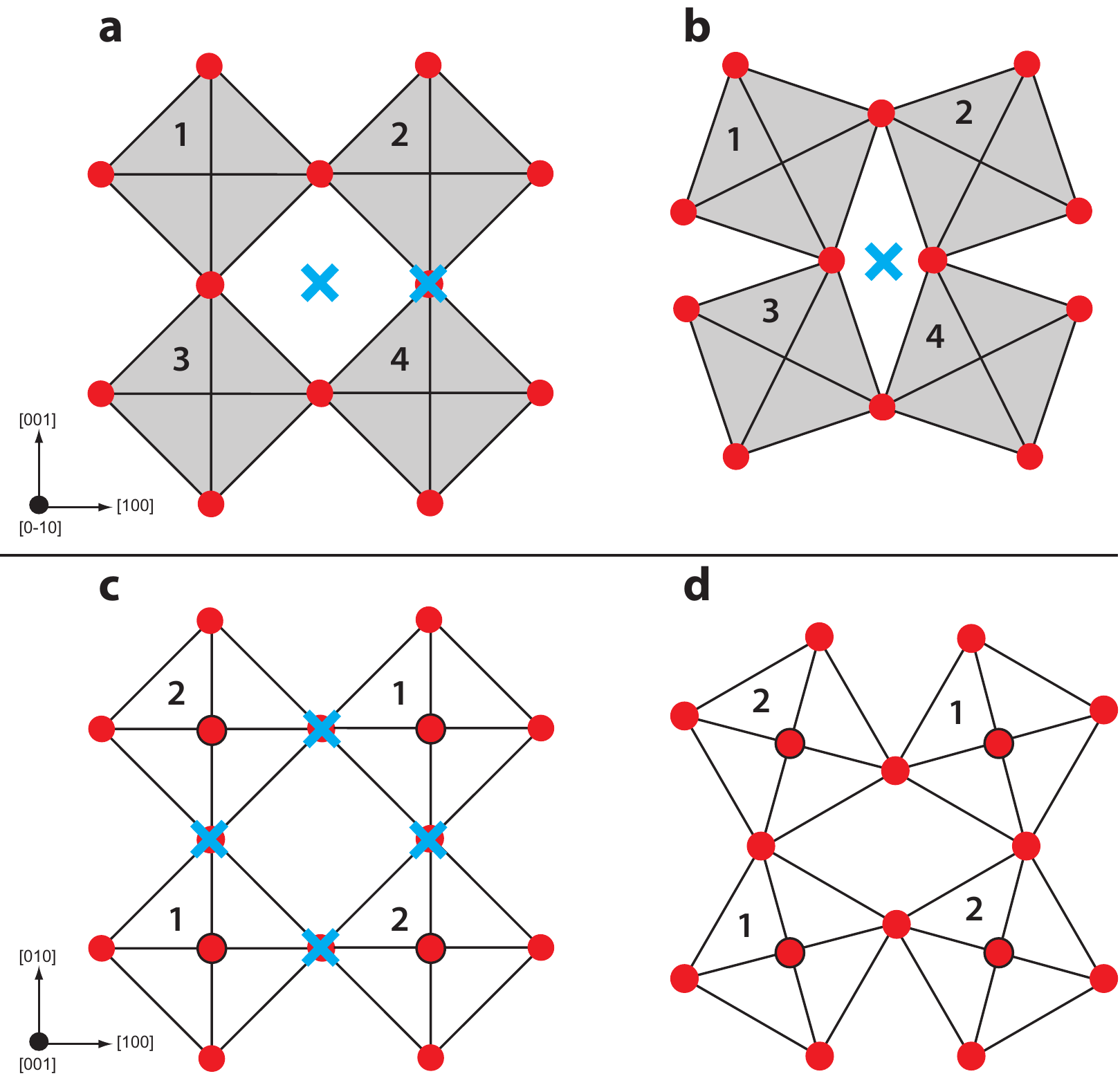}
\caption{\label{spins}\textbf{Effect of $X^-_3$ and $X^+_2$ on antiferromagnetic vector within a bilayer of Ca$_3$Mn$_2$O$_7$. } 
Schematic showing inversion centers, $\times$, with respect to the spins and antiferromagnetic vector $\bm{\mathrm{L}_0} = S_1 - S_2 - S_3 + S_4$ for \textbf{a} and \textbf{c}, undistorted octahedra ($I4/mmm$) and  \textbf{b}, $X^-_3$ ($Cmcm$) and \textbf{d}, $X^+_2$ ($Cmca$) distorted octahedra. Note in the bottom figure that the inversion center along [001] between, e.g., $S_1$ and $S_3$, is preserved. All numbers refer to the unique spins located in the center of the octahedra.}
\end{figure}

In the antiferromagnetically {\it  ordered} phase, we perform an energy expansion in the magnetization and in an order parameter, $\mathcal{P}$, that has similar transformational properties as the polarization about the saddle point along the $X^+_2$ switching path, i.e., at $Q_{X^-_3} \ne 0$ $\&$ $Q_{X^+_2} = 0$.
This corresponds to an orthorhombic  $Cmcm$ paraelectric reference structure, for which symmetry analysis and Figures~\ref{spins}a and b show the ordered antiferromagnetic vector $\bm{\mathrm{L}_0} = S_1-S_2-S_3+S_4$ being even with respect to space inversion, \textit{i.e.}, $\mathcal{I}${\bf L$_0$}  = $S_4-S_3-S_2+S_1$ = +{\bf L$_0$}. 
Note that we confirmed that $Cmcm$ is the saddle point structure by performing full structural relaxations from first principles within this $Cmcm$ tilt only structure  and  within the $Cmca$ rotation only ($Q_{X^-_3} = 0$ $\&$ $Q_{X^+_2} \ne 0$) structure. The tilt only structure is indeed closer in energy to the $A2_1am$ ground state.
Here, in this antiferromagnetic-paraelectric structure symmetry allows the magnetization, $M$, to couple to even powers of $\mathcal{P}$, where the lowest order coupling is of the form $$E_{int}=\gamma L_0 M,$$ where $\gamma$ is a coupling coefficient. A standard Landau approach [1] leads to
\begin{eqnarray}
M_0 &\sim&  \gamma L_0 + \mathcal{O}(\mathcal{P}_0^2) \\
\Delta_M &\sim & \gamma_2(L_0 \mathcal{P}_0) \Delta_{\mathcal{P}},
\end{eqnarray}
where $\Delta_{\mathcal{P}}$ and $\Delta_M$ are small changes about the equilibrium values of the magnetization ($M_0$) and $\mathcal{P}_0$ due to an applied electric field.
This analysis makes clear that switching the physical polarization with an electric-field to its symmetry-equivalent state by following the hybrid order parameter along the $X^+_2$ switching path does not switch the direction of the magnetization. The $X^+_2$ rotation mode, however, does induce the linear magnetoelectricity as seen in Eq$.2$.

Similar to the previous analysis, we perform an energy expansion in $\mathcal{P}$ and magnetization about the saddle point along the $X^-_3$ switching path, i.e., $Q_{X^-_3} = 0$ $\&$ $Q_{X^+_2} \ne 0$. This
corresponds to an orthorhombic  $Cmca$ paraelectric reference structure where symmetry analysis and Figures~\ref{spins}a, c, and d show the ordered antiferromagnetic vector $\bm{\mathrm{L}_0} = S_1-S_2-S_3+S_4$ is now odd with respect to space inversion, \textit{i.e.}, $\mathcal{I}${\bf L$_0$}  =  $S_3-S_4-S_1+S_2$ = -{\bf L$_0$}. We again confirmed that $Cmca$ is the saddle point structure by performing full structural relaxations.
Here, in this antiferromagnetic-paraelectric structure symmetry allows $M$ to couple to odd powers of $\mathcal{P}$, where the lowest order coupling is of the form $$E_{int}=\xi L_0 \mathcal{P} M,$$ where $\xi$ is a coupling coefficient. Again, using a standard Landau approach [1] leads to
\begin{eqnarray}
M_0& \sim& \xi L_0 \mathcal{P}_0 \\
\Delta_M &\sim& \xi L_0 \Delta_{\mathcal{P}}, 
\end{eqnarray}
where we now see that the equilibrium magnetization is proportional to the equilibrium polarization [2]. It vanishes at the paraelectric saddle point where $\mathcal{P}_0 = 0$, and is directly proportional to the strength of the ferroelectric distortion, i.e., $X^-_3$ induces weak-ferromagnetism.
Therefore switching the physical polarization with an electric-field to its symmetry-equivalent state by following the hybrid order parameter along the $X^-_3$ switching path does indeed switch the direction of the magnetization.

\textbf{References}\\
$[1]$ E.A. Turov, \textit{Physics-Uspekhi} \textbf{37}, 303 (1994).\\
$[2]$ D. L. Fox and J. F. Scott, 
{J. Phys. C: Solid State Phys.} {\bf 10}, L329 (1977).  


\


\begin{thebibliography}{32}
\bibitem[{\citenamefont{{In compressively strained films the direct transition
  may also be possible, although an alternative path as depicted in
  Figure~\ref{domains}b would be $I4/mmm \rightarrow Cmca \rightarrow
  A2_1am$}}()}]{zzzStrainSeq}
\bibinfo{author}{\bibnamefont{{In compressively strained films the direct
  transition may also be possible, although an alternative path as depicted in
  Figure~\ref{domains}b would be $I4/mmm \rightarrow Cmca \rightarrow
  A2_1am$}}}.
\end{thebibliography}
\end{document}